\documentclass[a4paper,fleqn,authoryear]{cas-dc}

\usepackage{graphicx}
\usepackage{algorithm}
\usepackage{algpseudocode}
\usepackage{xcolor}
\usepackage{listings}
\usepackage{amsfonts}
\usepackage{subcaption}

\newcommand{\review}[1]{\textcolor{black}{#1}}

\newcolumntype{L}[1]{>{\raggedright\let\newline\\\arraybackslash\hspace{0pt}}m{#1}}
\newcolumntype{C}[1]{>{\centering\let\newline\\\arraybackslash\hspace{0pt}}m{#1}}
\newcolumntype{R}[1]{>{\raggedleft\let\newline\\\arraybackslash\hspace{0pt}}m{#1}}
\definecolor{listinggray}{gray}{0.9}
\definecolor{lbcolor}{rgb}{0.9,0.9,0.9}
\usepackage{amsmath}
\usepackage[authoryear,longnamesfirst]{natbib}
\usepackage[super]{nth}
\usepackage{balance}
\usepackage{xurl}

\begin{document}
\let\WriteBookmarks\relax
\def\floatpagepagefraction{1}
\def\textpagefraction{.001}
 
\shorttitle{Malleability for Resource Optimization in HPC Workloads}    
\shortauthors{S. Iserte et al.}  
\title [mode = title]{Resource Optimization with MPI Process Malleability for Dynamic Workloads in HPC Clusters}

\author[1]{Sergio Iserte}[orcid=0000-0003-3654-7924]
\ead{sergio.iserte@bsc.es}

\author[2]{Iker Martín-Álvarez}[orcid=0000-0002-3337-3298]
\author[3]{Krzysztof Rojek}[orcid=0000-0002-2635-7345]
\author[2]{José I. Aliaga}[orcid=0000-0001-8469-764X]
\author[2]{Maribel Castillo}[orcid=0000-0002-2826-3086]
\author[3]{Weronika Folwarska}[orcid=0009-0002-9003-4898]
\author[1]{Antonio J. Peña}[orcid=0000-0002-3575-4617]

\affiliation[1]{organization={Barcelona Supercomputing Center (BSC)}, 
            city={Barcelona},
            country={Spain}
            }
            
\affiliation[2]{organization={Universitat Jaume I (UJI)},
            city={Castelló de la Plana},
            country={Spain}}

\affiliation[3]{organization={Department of Computer Science, Częstochowa University of Technology (PCZ)},
            city={Częstochowa},
            country={Poland}}

\begin{abstract}
Dynamic resource management is essential for optimizing computational efficiency in modern high-performance computing (HPC) environments, particularly as systems scale. 
While research has demonstrated the benefits of malleability in resource management systems (RMS), the adoption of such techniques in production environments remains limited due to challenges in standardization, interoperability, and usability. 
Addressing these gaps, this paper extends our prior work on the Dynamic Management of Resources (DMR) framework, which provides a modular and user-friendly approach to dynamic resource allocation.
Building upon the original DMRlib reconfiguration runtime, this work integrates new methodology from the Malleability Module (MaM) of the Proteo framework, further enhancing reconfiguration capabilities with new spawning strategies and data redistribution methods. 
In this paper, we explore new malleability strategies in HPC dynamic workloads, such as merging MPI communicators and asynchronous reconfigurations, which offer new opportunities for dramatically reducing memory overhead.
The proposed enhancements are rigorously evaluated on a world-class supercomputer, demonstrating improved resource utilization and workload efficiency. Results show that dynamic resource management can reduce the workload completion time by $40\%$ and increase the resource utilization by over $20\%$, compared to static resource allocation.
\end{abstract}


\maketitle

\section{Introduction}\label{intro}
In the evolving landscape of high-performance computing (HPC), dynamic resource management has become increasingly critical for efficiently leveraging the vast computational capabilities required by modern scientific applications. These range across domains such as aerodynamics~\cite{SanderseReviewAerogenerators}, industrial mixing~\cite{CFDMixing}, nuclear reactor safety~\cite{CFDNucelar}, and wastewater treatment~\cite{CFDAeration}, among others.

The advent of Exascale computing, marked by unprecedented levels of parallelism and architectural complexity—as evidenced in the TOP500 list~\cite{Dongarra2011}—has amplified the demand for flexible and adaptive resource management strategies. In response, initiatives like the EuroHPC Joint Undertaking (JU)~\cite{eurohpc_ju} have supported numerous European projects dedicated to developing dynamic resource managers tailored to domain-specific requirements. As the community seeks to maximize scientific productivity and resource efficiency, dynamic resource management has emerged as a vital strategy for modern HPC systems.

\review{
Several studies~\cite{Iserte2020, Chadha2021} have shown that dynamic jobs can improve critical metrics such as resource utilization, makespan, and job turnaround time. While moldability offers flexibility at job launch, malleability—the ability to adapt resource usage during execution—enables continuous optimization, further enhancing system efficiency. Beyond scheduling, malleability can be leveraged at the file system level~\cite{Sanchez2023} to improve I/O performance, and has also been used to reduce energy consumption in HPC workloads~\cite{Iserte2019a, Cascajo2023}.
}

Despite these advances, integrating dynamic resource management into production environments remains challenging due to limitations in standardization, interoperability, and usability. Many existing solutions lack the modularity needed to support diverse runtimes and impose substantial adaptation burdens on applications. Addressing these barriers requires a unified framework that combines flexible resource allocations with comprehensive malleability support at both application and system levels.

This paper builds upon and significantly extends our previous work on the Dynamic Management of Resources (DMR) framework~\cite{iserte_towards_2025}, which was developed to democratize dynamic resource management via a modular and user-friendly application programming interface (API). The original DMR framework introduced a stacked architecture capable of integrating multiple reconfiguration managers, and offered an API that abstracts away low-level details of reconfiguration runtimes and resource managers.

A key outcome of that work was the seamless integration of DMRlib~\cite{Iserte2020}, our reconfiguration runtime, with MaM, the malleability engine from the Proteo framework~\cite{proteo_2024}. This integration enabled a comprehensive suite of spawning strategies and data redistribution techniques suitable for production environments. We demonstrated its feasibility through a prototype evaluation using a malleable implementation of the Multidimensional Positive Definite Advection Transport Algorithm (MPDATA)~\cite{SMO2006} in a controlled testbed.

This paper presents a substantially extended version of that work. We evaluate the enhanced DMR framework at scale on a world-class supercomputing facility and demonstrate how the new capabilities lead to improved resource utilization, coding usability, and scientific productivity. Furthermore, we explore novel strategies for malleability in MPI workloads —including asynchronous reconfigurations and communicator merging— that significantly reduce memory overhead and increase flexibility.

In particular, this paper introduces the following key contributions: 
\begin{enumerate} 
\item An improved MaM module capable of dynamically selecting the optimal reconfiguration strategy. 
\item An enhanced MPDATA implementation with a redesigned data layout to fully exploit MaM’s redistribution capabilities. 
\item A new plugin for malleability-aware Slurm, supporting reconfiguration to any natural number of resources. 
\item A detailed experimental evaluation of different reconfiguration strategies on a large-scale HPC system, demonstrating the importance of optimal resource management. 
\end{enumerate}

The rest of the paper is organized as follows: Section~\ref{sec:background} provides an overview of the technologies used; Section~\ref{sec:related} discusses related work on MPI malleability and dynamic resource management; Section~\ref{sec:res} details the architectural division between DMR and DMRlib and the integration of MaM, with an emphasis on new features that improve resource usage; Section~\ref{sec:experiments} presents and analyzes experimental results; 
Section~\ref{sec:app} discusses the applicability of the presented malleability framework, and finally, Section~\ref{sec:conclusions} concludes with a summary.

\section{Background}\label{sec:background}
This section highlights the main technologies involved in this research, namely: 1) the DMRlib reconfiguration framework, 2) a malleability-enabled implementation of Slurm, 3) the Malleability Module (MaM) of Proteo, and 4) the scientific application MPDATA, with which we evaluate the presented malleability ecosystem.

\subsection{DMRlib}\label{subsec:dmrlib}
The Dynamic Management of Resources Library (DMRlib)~\cite{Iserte2020} is an advanced API that facilitates the integration of malleability into HPC applications. It provides a communication interface between the parallel distributed runtime (PDR) and the resource management system (RMS), allowing users to transparently manage processes and resources, thereby improving productivity and optimizing resource utilization in HPC environments.

Fundamentally, DMRlib provides passive dynamic resource management through the following steps:
\begin{itemize}
\item During job execution, DMRlib periodically communicates with the RMS to signal jobs' readiness for reconfiguration.
\item This interaction occurs at a designated synchronization point within the code, triggering the reconfiguration process. In iterative applications, the end of an iteration frequently serves as an ideal synchronization point.
\item The RMS then evaluates the overall system status to decide whether a reconfiguration is required and sends this decision back to DMRlib.
\item If this communication leads to a change in job size, the RMS reallocates resources accordingly, which may involve either increasing or decreasing the number of processes.
\item In the end, according to the application's guidelines, the data is redistributed across the processes, and the job continues execution with the updated process configuration, beginning precisely from the reconfiguration point.
\end{itemize}

DMRlib has proven to be a highly effective and versatile malleability tool, with its impact demonstrated by a wide range of published success stories and implementations that continue to demonstrate its ability to improve the efficiency and flexibility of HPC applications in diverse environments~\cite{Iserte2019a, Iserte2020, iserte_towards_2025}.

\subsection{Malleability-enabled Slurm}\label{subsec:slurm}
Slurm is a highly robust and scalable open-source cluster manager and job scheduling system specifically designed for Linux-based clusters~\cite{Yoo2003}. Slurm effectively handles large-scale distributed systems, making it an excellent solution for HPC environments. Its modular architecture allows for flexible configuration and extension, while it supports various advanced features, including job prioritization, resource allocation, and job dependencies. Additionally, Slurm can manage workloads efficiently by providing fine-grained control over job execution and resource utilization, with the ability to schedule both parallel and sequential jobs. 

The basic deployment of Slurm involves several components that work together to manage and schedule jobs efficiently in an HPC environment. At the core, there is a daemon called \textit{slurmd}, which runs on each compute node. This daemon is responsible for managing the execution of jobs on that node, including the allocation of resources, monitoring job status, and reporting back to the central system. The \textit{slurmctld} is the central daemon that runs on the management node. It oversees the entire scheduling process, including the assignment of jobs to compute nodes, prioritization of jobs, and handling of resource allocation policies.


The daemons are responsible for overseeing the management of key elements within the system, including nodes, queues, allocations and job steps (sets of tasks within a job). By efficiently coordinating these components, they ensure optimal resource utilization and the proper execution of jobs within the Slurm-managed cluster environment. The available nodes within a partition are assigned to jobs in the priority queue.

DMRlib includes a specialized version of Slurm that supports malleable jobs, enabling dynamic resource management during job execution. This functionality is facilitated by a custom resource selection plugin that links DMRlib with Slurm, allowing for seamless reallocation of resources as needed.

Slurm monitors the allocation of resources to jobs during a resize operation, allowing for adjustments in job allocations across nodes.
On the one hand, $\textit{Job}\ A$ could be expanded as follows:
 \begin{enumerate}
  \item Submit a new $\textit{Job}\ B$ with dependency to the initial $\textit{Job}\ A$. $\textit{Job}\ B$ requests the number of nodes (NB) to be added to the allocated nodes (NA) by $\textit{Job}\ A$.
  \item Update $\textit{Job}\ B$, setting its number of nodes to 0.
  This produces a set of NB nodes not allocated to any job.
  \item Cancel $\textit{Job}\ B$.
  \item Update $\textit{Job}\ A$ and set its number of nodes to NA+NB.
 \end{enumerate}
On the other hand, $\textit{Job}\ A$ could be shrunk by updating $\textit{Job}\ A$'s number of nodes to the desired size.

After these steps, the environment variables for $\textit{Job}\ A$ in Slurm are updated accordingly. It is important to note that these commands do not impact the current status of the running job. The user continues to be responsible for managing any malleability process and for redistributing the data as needed.

\subsection{Proteo}\label{subsec:background-proteo}
Proteo is a highly configurable framework designed to facilitate the creation of benchmarks and experimental environments to study the effects and integration of malleability in real-world HPC applications~\cite{proteo_2024}. It is composed of various modules that enable users to incorporate malleability to MPI-based applications, implement malleable versions, and monitor performance malleability  metrics.

In this work, we specifically leverage the Malleability Module (MaM) of Proteo to enhance the malleability capabilities of DMRlib. MaM acts as the malleability engine and its novel API is adopted by  DMRlib to provide advanced spawning strategies and data redistribution techniques. This integration offers increased flexibility and efficiency in adapting resource allocations to the dynamic needs of applications.

MaM is responsible for managing application reconfigurations by adjusting the number of active processes at runtime. Its functionality is centered around two primary tasks: \textit{process management} and \textit{data redistribution}.

In the process management stage, MaM assumes an initial group of $NS$ (source) processes and a target size of $NT$ (target) processes. Two main methods are supported:
\begin{itemize}
    \item The \textit{Baseline} method always spawns $NT$ new processes, regardless of the existing group, with the connection between old and new processes handled via either intra- or inter-communicators.
    \item The \textit{Merge} method is more efficient, as it retains the minimum of $NS$ and $NT$ existing processes and spawns only the additional processes needed to reach $NT$. When shrinking, it removes the excess processes without spawning new ones. Unlike Baseline, Merge exclusively relies on intra-communicators, which are generally more efficient for tightly coupled applications.
\end{itemize}

MaM also supports multiple optimizations to further enhance reconfigurations:
\begin{itemize}
    \item The \textit{Asynchronous} strategy offloads spawning operations to auxiliary threads, allowing the application to continue execution while new processes are being launched in the background.
    \item The \textit{Parallel} strategy distributes the creation of the new \texttt{MPI\_COMM\_WORLD} communicator across different nodes in parallel, minimizing synchronization bottlenecks. This is particularly beneficial when aiming for future shrink operations, as it simplifies intra-node reductions.
\end{itemize}

In the data redistribution phase, MaM handles the migration of application data from the original communicator to the newly created one. This transfer is performed efficiently, with support for both \textit{synchronous} communication of variable data and \textit{asynchronous} communication of constant data, depending on the application's characteristics and performance goals.

\subsection{MPDATA Algorithm}
MPDATA is leveraged to generate the workload with which we evaluate the presented framework. 
The application is based on a 3D implementation of the Multidimensional Positive Definite Advection Transport Algorithm (MPDATA)~\cite{SMO2006}, which serves as a representative example of a real-world stencil computation. 
The goal is to highlight key computational and memory access patterns relevant to the optimization strategies discussed in the subsequent sections.

Notably, the MPDATA algorithm forms the foundation of the multiscale fluid solver EULAG (Eulerian/semi-Lagrangian fluid solver) \cite{ROJ17a}. Its primary function is to compute the advection of a nondiffusive scalar field $\Psi$ within a given velocity field, represented by the continuity equation:

\begin{equation}
{{\partial \Psi} \over {\partial t}} + \text{div} ( \mathbf{V} \Psi ) = 0 ,
\end{equation}
where $\mathbf{V}$ denotes the velocity vector. The algorithm utilizes finite difference methods for spatial discretization~\cite{ROJ17b}.

MPDATA operates iteratively and is designed for rapid convergence. In the initial step, the advection of a prognostic field $\psi$ is computed using the donor-cell approximation, providing first-order accuracy~\cite{ROJ17b}. Subsequent iterations introduce corrections that elevate the accuracy to second order in both space and time. These corrections involve reapplying the donor-cell approach, but now incorporating anti-diffusive velocities derived from the computed fields.


As a forward--in--time algorithm, MPDATA advances calculations in discrete time steps, with the number of iterations determined by the specific physical scenario being modeled. Each iteration involves five input arrays and produces a single output array, which serves as input for the subsequent time step.

\review{
To achieve scalability and efficiency, a hybrid parallelization approach combining a distributed memory model for inter-node communication and a shared memory model for intra-node computations is adopted. In the distributed memory model, the computational domain is decomposed along the first $X$ dimension and distributed across nodes. Each decomposed part (chunk) of size $N \times (M+1) \times (L+1)$ exchanges halo areas of size $3 \times (M+1) \times (L+1)$ with neighboring nodes using MPI, ensuring continuity of calculations across the global domain. Here, $M$, $N$, and $L$ denote the dimensions of the 3D computational grid used in the MPDATA algorithm, corresponding respectively to the number of grid points in the $X$, $Y$, and $Z$ directions.
}

\begin{figure*}[tbp]
    \centering
    \includegraphics[width=0.65\linewidth]{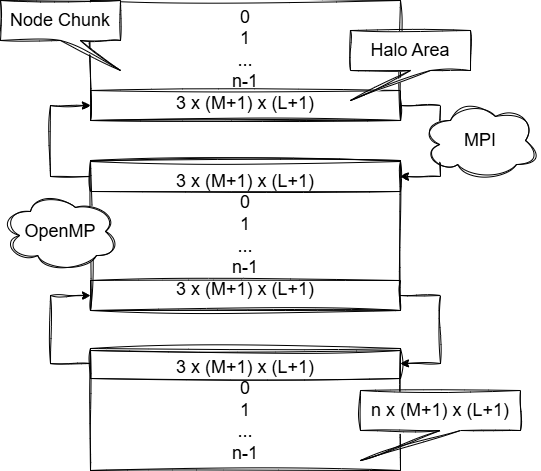}
    \caption{Scheme of the MPDATA data redistribution}
    \label{fig:MPDATA-scheme}
\end{figure*}

Within each node, the shared memory model leverages OpenMP to parallelize computations across the available CPU cores. By benefiting from the shared memory architecture within nodes, we reduce the communication overhead compared to inter-node exchanges. The data distribution scheme is shown in Figure~\ref{fig:MPDATA-scheme}.

\section{Related Work}\label{sec:related}
This section explores on--the--fly reconfiguration solutions integrated into production-grade resource managers, emphasizing the need for a modular framework to support diverse malleability engines from a system-wide perspective.

We highlight the most prominent MPI-based solutions for dynamic processes and job malleability. For further details, \cite{aliaga_survey_2022} offers a comprehensive survey on spawn-based process malleability, while \cite{tarraf_malleability_2024} provides a broader overview of the state of the art in dynamic resource management.

ReSHAPE~\cite{Sudarsan2007} provides an integrated framework for adapting to varying workloads, combining specialized reconfiguration libraries, a scheduler, and a runtime system. However, the tight coupling of these components requires applications to be specifically tailored for compatibility with ReSHAPE.

The Power-Aware Resource Manager (PARM)~\cite{Sarood2014} leverages over-provisioning, power capping, and job malleability to optimize throughput under strict power constraints in over-provisioned facilities. Its malleability depends on Charm++ runtime support, which dynamically redistributes computational tasks across processors.

The approach in~\cite{Prabhakaran2015} integrates Adaptive Message Passing Interface (AMPI)~\cite{AMPI06}, a Charm++-based MPI implementation, with the Torque/Maui job scheduler to address malleable jobs. This integration establishes a communication interface between the Charm++ runtime and Torque/Maui.

Elastic MPI~\cite{compres_infrastructure_2016} provides the infrastructure and API extensions for running malleable MPI applications, utilizing Slurm and MPICH. Applications using this API can periodically detect reconfiguration events triggered by Slurm. MPICH is also extended with new functions to replace or enhance its standard implementation. However, this approach is limited to MPICH, lacks support for other MPI implementations, and does not handle data redistribution.

\review{
The authors in~\cite{bhattarai_dynamic_2024} present a combination of workflow manager and resource management system that can accommodate a fine-grain elastic resource allocation during the execution of a workflow based on the Parsl workflow manager and PMIx-enabled SLURM.
}

Thanks to these advancements, malleability is becoming an increasingly accessible technology. As expected, these solutions introduce overhead during data redistribution, whether by loading data from disk or transferring it over the network. Nevertheless, the scheduling overhead in solutions that integrate with resource management systems (RMS) is reported to be negligible.

Originally, DMRlib was built using Nanos++ (the OmpSs runtime) and Slurm as PDR and RMS, respectively. In this work, the DMRlib design has been refactored to support various PDR and RMS systems. Specifically, we introduce the Dynamic Resource Manager (DMR) as a universal programming layer for dynamic resources, with DMRlib acting as the communication layer between the application, the PDR, and the RMS.
Particularly, the original work~\cite{iserte_towards_2025} introduced in the new modular design the support for standard MPI, the most common distributed parallel programming paradigm in supercomputers, as the PDR of the framework.

Beyond spawn-based solutions, modern approaches to dynamic process management are emerging, leveraging MPI Process Sets (PSets)—a key concept introduced in MPI~4.0 under the MPI Sessions model. Although implementation varies across MPI distributions, efforts such as ParaStation MPI~\cite{partec_docs} and Dynamic Processes with PSets (DPP)~\cite{huber_towards_2022} are expanding their applicability in the context of dynamic resources. \review{In this space, DMR provides seamless support for DPP~\cite{huber_bridging_2025} and it is ready to enable users to run applications across different malleability backends with the same frontend code}.


\section{Optimizing Resource Utilization}\label{sec:res}
This section outlines the implemented features at each level: malleability framework, malleability engine, and malleable application.

\subsection{Dynamic Resource Management}\label{subsec:dynres}
Originally, DMRlib was introduced as a framework for managing dynamic resources within HPC applications, offering a user-friendly syntax to ease the development of malleable codes (see Section~\ref{subsec:dmrlib}). However, the renewed interest in dynamic resource management has led several institutions to develop their own frameworks targeting similar goals. 

\review{In the absence of standardized malleability support in MPI, DMRlib fills this gap by offering backend-agnostic support for various reconfiguration runtimes. This abstraction allows both application source code and end users to remain unaware of the specific mechanisms managing resources and processes behind the scenes, significantly lowering the barrier to adopting dynamic execution models.}

\begin{figure*}[htb]
    \centering
    \includegraphics[width=0.95\linewidth]{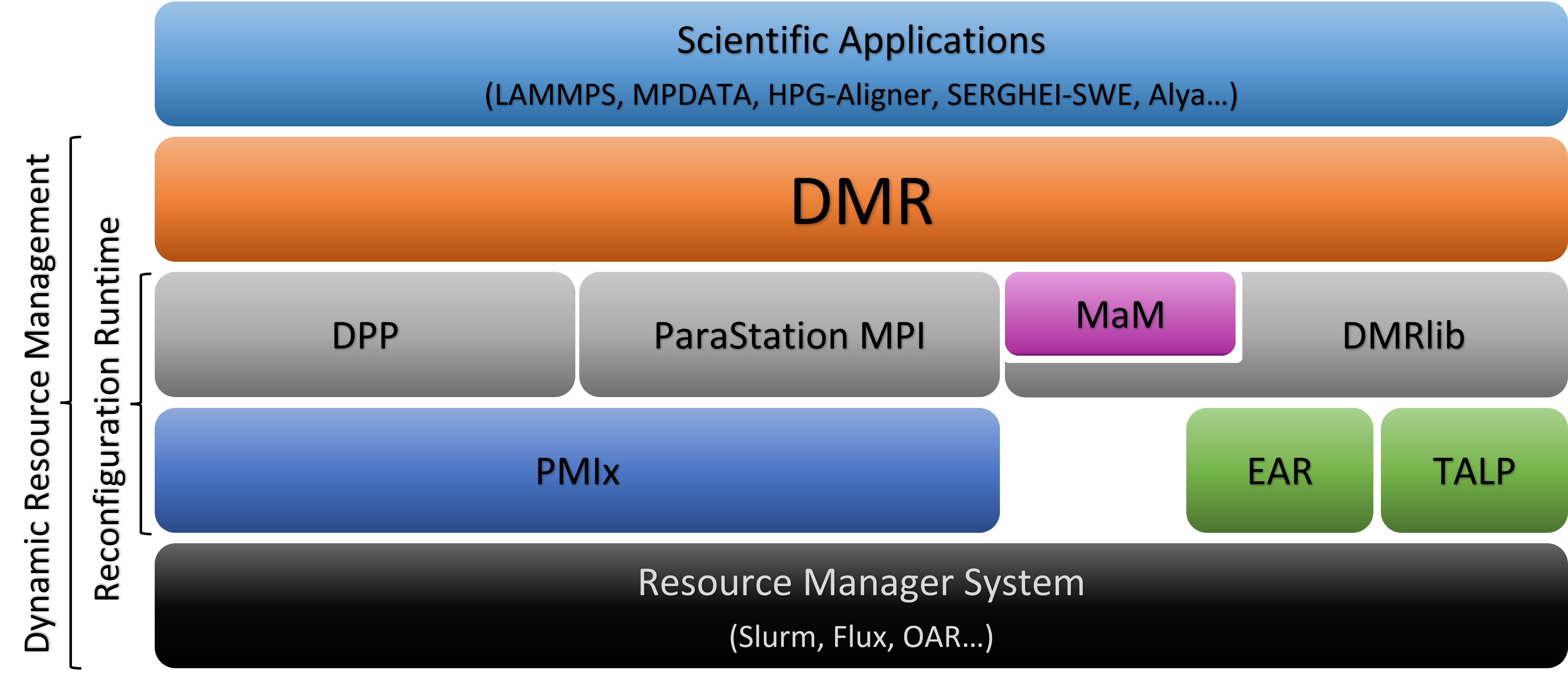}
    \caption{Dynamic Management of Resources Software Stack}
    \label{fig:dmr-layers}
\end{figure*}

In particular, an excision of DMRlib is created to provide a user API independent of the underlying dynamic resource manager. This excision is illustrated in Figure~\ref{fig:dmr-layers} in the layer ``DMR'', which is responsible for hiding the dynamic resources frameworks from the users.

DMR is extended with an interoperability interface for the new version of DMRlib, as it was done for DPP~\cite{huber_bridging_2025}.
Besides, DMRlib is expanded with a pure-MPI backend able to operate without OmpSs, which was a strong dependency in the previous version~\cite{Iserte2020}. Although the logic is similar, DMRlib, alternatively, substitutes OmpSs pragma directives and Nanos++ runtime operations in favor of MPI standard functions, allowing the DMR interface for DMRlib to rely solely on MPI.

In this regard, scientific applications can enable dynamicity by leveraging the new DMR API. Listing~\ref{code:dmr-macros} contains the three mandatory interfaces to include in the user code.
These wrappers expect a series of user functions to perform the operations during the execution initialization and finalization (the first and third in the listing, respectively) and the data redistribution functions (the second) in the case of a resize.

\begin{lstlisting}[float, caption=DMR interfaces., label=code:dmr-macros, captionpos=b, numbers=none, framexleftmargin=0pt, xleftmargin=0cm]
DMR_INIT( initialize(...), redistribution(...) );
DMR_RECONFIGURE( redistribution(...) );
DMR_FINALIZE( finalize(...) ); 
\end{lstlisting}

While \texttt{DMR\_INITIALIZE} and \texttt{DMR\_FINALIZE} can be invoked at the beginning and the end of the code, respectively, 
\texttt{DMR\_RECONFIGURATION}
is called whenever the execution is ready to be reconfigured, in other words, at synchronization points. 
For instance, iterative applications, such as multiphysics simulations~\cite{VAZQUEZ201615}, present a main loop in which each iteration corresponds to a complete time step where the processes are synchronized previously to continue with the next time step. 
That structure can be seen  as a periodic reconfiguration point to trigger malleability.

At that point, if the RMS determines to reconfigure the job, data must be redistributed accordingly to the user function ``redistribution''. For this purpose, this function is called when target processes have to set their data according to the current status of the execution. These data come from the parent processes sent during the reconfiguration stage.

Following, the three basic DMR interfaces are introduced:
\begin{itemize}
    \item \texttt{DMR\_INITIALIZE}: Listing~\ref{code:dmrlib-ini} shows the DMR interface for DMRlib which discriminates between ranks initializing from scratch or receiving the state from former ranks.
\begin{lstlisting}[float, caption=\textit{DMR\_INITIALIZE} interface for DMRlib., label=code:dmrlib-ini, captionpos=b]
#define DMR_INIT(initialize, redistribution) {    \      
    MPI_Comm_size(MPI_COMM_WORLD, &DMR_comm_size);\
    MPI_Comm_rank(MPI_COMM_WORLD, &DMR_comm_rank);\
    MPI_Comm_get_parent(&DMR_INTERCOMM);          \
    if (DMR_INTERCOMM != MPI_COMM_NULL) {         \
        redistribution;                           \
        MPI_Comm_disconnect(&DMR_INTERCOMM);      \
    } else {                                      \
        initialize;                               \
    }                                             \
}
\end{lstlisting}
    \item \texttt{DMR\_RECONFIGURE}: Listing~\ref{code:dmrlib-rec} contains the DMR interface for DMRlib, which in the case of a non-inhibited reconfiguration,  determines the action and actuates accordingly, executing the data redistribution and detaching communicators before continuing.
\begin{lstlisting}[float, caption=\textit{DMR\_RECONFIGURE} interface for DMRlib., label=code:dmrlib-rec, captionpos=b]
#define DMR_RECONFIGURATION(redistribution) {    \
    if (!_inhibit_iter()){                       \
        if (DMR_Reconfiguration( ... )) {        \
            redistribution;                      \
            DMR_Detach( ... );                   \
        }                                        \
    }                                            \
}
\end{lstlisting}
    \item \texttt{DMR\_FINALIZE}: Listing~\ref{code:dmrlib-fin} illustrates the DMR interface for DMRlib which executes the user function ``finalize''.
\begin{lstlisting}[float, caption=\textit{DMR\_FINALIZE} interface for DMRlib., label=code:dmrlib-fin, captionpos=b]
#define DMR_FINALIZE(finalize) {                \
    finalize;                                   \
}
\end{lstlisting}
\end{itemize}

To determine the reconfiguration action, this work leverages a Slurm resource selection plugin that supports malleability~\cite{Iserte2020}. 
When a malleable job triggers a reconfiguration, Slurm decides which action to take: \textit{nothing} to continue with the current configuration, or \textit{expand} or \textit{shrink} to a specific number of processes.

The Slurm plugin presented in this paper, \texttt{select/natural}, is an evolution of the plugin based on \texttt{select/power2}~\cite{iserte_towards_2025}, which in turn was based on \texttt{select/linear}.
Algorithm~\ref{alg:slurm} depicts how \texttt{select/natural} determines the resultant number of nodes assigned to the job after a reconfiguration, if applicable. 
As its name suggests, \texttt{select/natural} allows reconfigurations to or from any natural number of nodes, offering greater flexibility compared to its predecessor, \texttt{select/power2}, which restricts resizes within only powers of two values.

This policy aims to maximize the resource utilization rate. 
Specifically, the algorithm first checks if there are no pending jobs in the queue. 
If none exist, it expands the triggering job to its maximum possible size, provided resources are available. 
If there are pending jobs, the algorithm evaluates them to determine if one can be initiated using part of the resources currently allocated to the triggering malleable job. 
If feasible, the necessary nodes are released from the triggering job (never exceeding its given minimum size), and the selected pending job is given the highest priority to be initiated. If no awaiting job can be started, the algorithm attempts to expand the triggering job, if resources permit.

\begin{algorithm}[htb]
\small
\caption{Reconfiguration policy implemented in the Slurm plugin \texttt{select/natural}}\label{alg:slurm}
\begin{algorithmic}
\State $resultant\_nodes \gets 0$
\If{\textbf{not} $pending\_jobs()$}
    \If{$available\_resources()$}
         \State $resultant\_nodes \gets expand\_to()$
    \EndIf
\Else 
    \If{$job\_can\_be\_initiated\_with\_my\_resources()$}
        \State increase\_selected\_pending\_job\_priority()
        \State $resultant\_nodes \gets shrink\_to()$
    \Else 
        \If{$available\_resources()$}
            \State $resultant\_nodes \gets expand\_to()$
        \EndIf
    \EndIf
\EndIf
\State \Return $resultant\_nodes$
\end{algorithmic}
\end{algorithm}

\subsection{Reconfiguration Engine}\label{subsec:reconf}
DMRlib is enhanced with MaM's robust process spawning and data redistribution capabilities. In this work, MaM replaces the original reconfiguration engine in DMRlib, as illustrated in Figure~\ref{fig:dmr-layers}.
In this regard, DMRlib leverages MaM features without interfering with other  functionalities such as the communication with Slurm, as well as performance-aware (via TALP~\cite{talp}) and energy-aware (via EAR~\cite{ear}) reconfiguration policies.

For example and analogously to the DMRlib interface presented in listings~\ref{code:dmrlib-ini},~\ref{code:dmrlib-rec}, and~\ref{code:dmrlib-fin}, listings~\ref{code:proteo-ini},~\ref{code:proteo-rec}, and~\ref{code:proteo-fin} contain the respective DMR interfaces for MaM. These are the three interfaces:

\begin{itemize}
    \item \texttt{DMR\_INITIALIZE}: Listing~\ref{code:proteo-ini} presents the DMR interface for MaM, where the initialization tasks are performed. First, the MaM resize engine is initialized (line~2), determining if the processes pertain to the source or target group. For the source ranks, DMR initializes its basic data structures and registers in MaM the data that should be redistributed upon future resizes (line~11). Conversely, if the ranks are in the target group, each retrieves its data (line~6), and the root process (line~7) waits for the Slurm confirmation to ensure that the reconfiguration has been completed before proceeding (line~8).
\begin{lstlisting}[float, caption=\textit{DMR\_INITIALIZE} interface for MaM., label=code:proteo-ini, captionpos=b]
#define DMR_INIT(initialize, redistribution) {  \
    MaM_is_children = MAM_Init(&DMR_COMM);      \
    MPI_Comm_size(DMR_COMM, &DMR_comm_size);    \
    MPI_Comm_rank(DMR_COMM, &DMR_comm_rank);    \
    if (MaM_is_children) {                      \
        DMR_Retrieve_data(redistribution, ... );\
        if (DMR_comm_rank == 0) {               \
            DMR_Wait_stabilization( ... );      \
        }                                       \
    } else {                                    \
        DMR_Set_basic_data(initialize, ... );   \
    }                                           \
}
\end{lstlisting}
    \item \texttt{DMR\_RECONFIGURE}: Listing~\ref{code:proteo-rec} illustrates the DMR interface for MaM during reconfiguration. In this case, DMR first communicates with Slurm to determine whether a resize operation (global variable \textit{DMR\_\-action}) should occur (line~3). If a resize is required, MaM dynamically adjusts the application to the new resource allocation (line~6). Since MaM supports background resizing, the function may be called multiple times until the process is fully completed. Once the resize finishes, DMR returns the unneeded resources and notifies MaM to terminate any excess processes (line~8). Then, the data redistribution is concluded according to the resize action (line~9).
\begin{lstlisting}[float, caption=\textit{DMR\_RECONFIGURE} interface for MaM., label=code:proteo-rec, captionpos=b]
#define DMR_RECONFIGURATION(redistribution) {   \
    if (!_inhibit_iter() && DMR_action < 1) {   \
        DMR_action = DMR_Reconfiguration( ... );\
    }                                           \
    if (DMR_action >= 1) {                      \
        MAM_Checkpoint(&MaM_state, MAM_CHECK_COMPLETION);                \
        if (MaM_state == MAM_COMPLETED) {       \
            DMR_Detach( ... );                  \
            redistribution;                     \
        }                                       \
    }                                           \
}
\end{lstlisting}
    \item \texttt{DMR\_FINALIZE}: Listing~\ref{code:proteo-fin} contains the DMR interface for MaM. During the finalization, the interface ensures that non-blocking reconfigurations are terminated (line~2) and cleans up the memory (lines~4 and~5). 
\begin{lstlisting}[float, caption=\textit{DMR\_FINALIZE} interface for MaM., label=code:proteo-fin, captionpos=b]
#define DMR_FINALIZE(finalize) {                \
    if (MaM_state == MAM_PENDING)               \
        MAM_Checkpoint(&MaM_state, MAM_WAIT_COMPLETION);                 \
    finalize;                                   \
    MAM_Finalize();                             \
}
\end{lstlisting}
\end{itemize}

Notice that MaM has been extended to dynamically select the most suitable resizing technique of its catalog~\cite{iker_spawn, iker_redistribution}. 
Algorithm~\ref{alg:mam} describes the procedure created to determine which strategy minimizes the cost of reconfiguring.
By default, the algorithm selects the \textit{Parallel} strategy for process spawning and the \textit{Collective} method for data redistribution. 
In addition, the \textit{Merge} method is set as the default choice for both expansions and shrinkages. 
An exception is made in shrinkage scenarios where the number of target processes is smaller than the number of original ranks (i.e., the size of the communicator at job startup); in such cases, the \textit{Baseline} method with intra-communicators is preferred.

\begin{algorithm}[htb]
\small
\caption{Procedure to configure MaM parameters.}\label{alg:mam}
\begin{algorithmic}
\If{$user\_has\_custom\_config()$}
    \State \Return
\EndIf

\State $spawn\_strat \gets MAM\_SPAWN\_PARALLEL$
\State $redist\_method \gets MAM\_RED\_COLLECTIVE$
\If{$is\_expanding()$}
    \State $spawn\_method \gets MAM\_SPAWN\_MERGE$
\Else 
    \State $global\_original\_ranks \gets check\_global\_internodes\_qty()$
    \State $numT \gets get\_number\_targets()$
    \If{$numT < global\_original\_ranks$}
        \State $spawn\_method \gets MAM\_SPAWN\_BASELINE$
        \State $spawn\_strat \gets MAM\_SPAWN\_INTRACOMM$
    \Else 
        \State $spawn\_method \gets MAM\_SPAWN\_MERGE$
    \EndIf
\EndIf
\State \Return
\end{algorithmic}
\end{algorithm}

\subsection{Malleable Application}\label{subsec:app}
MPDATA has been extended to support malleability.
Listing~\ref{code:mpdata-dmr} showcases the \textit{main} function of the malleable version of MPDATA and how DMR API and MaM are leveraged. 

The listing illustrates an iterative MPI application, incorporating the necessary logic to enable dynamic resource management with DMR.
In particular, line~4 initializes the environment. 
Line~5 defines the malleability limits, specifying the lower and upper bounds.
Meanwhile, line~6 introduces an inhibitor to prevent consecutive reconfiguration checks.
The "MIN" and "MAX" values are detailed in Table~\ref{tab:jobs_conf} in Section~\ref{sec:experiments}.
At the end of each iteration, unless inhibited, a reconfiguration is evaluated (line~10). 
If a resize is scheduled, the operation described in Listing~\ref{code:proteo-rec} is performed.
Finally, the program releases memory and terminates.

\begin{lstlisting}[float, caption=MPDATA main function., label=code:mpdata-dmr, captionpos=b]
void main(int argc, char **argv) {
  MPI_Init_thread(&argc, &argv, MPI_THREAD_MULTIPLE, NULL);
  init();                       // initialization
  DMR_INIT(mam_set(), mam_retrieve());
  DMR_Set_parameters(MIN, MAX); 
  DMR_Inhibit_iter(INH_ITERS);
  while (DMR_it < TIME_STEPS) { // main loop
    update();                   // halo exchange
    mpdata3d();                 // computation
    DMR_RECONFIGURE(mam_retrieve(), release());
  }
  DMR_FINALIZE(release());      // finalize
  MPI_Finalize();
}
\end{lstlisting}

MaM implements predefined data redistribution patterns.
For this purpose, MaM can redistribute 1D data with just a few hints from the user. 
Listing~\ref{code:dmr-proteo-redist} shows the functions in MPDATA to indicate which data will be redistributed.
Furthermore, \texttt{mam\_set} is invoked once by the \textit{source} processes to declare the data pointers within the MaM runtime. This function requires an MPI derived datatype, created in \texttt{create\_slice\_datatype}, representing slices along the $N$ dimension. Each slice is treated as a single data unit containing all the values across the $M$ and $L$ dimensions. As a result, data can be represented as a flattened array in the $N$ dimension.
Finally, \texttt{mam\_retrieve} is called by the \textit{target} processes to obtain the new pointers of the redistributed data.
It is important to note that these functions do not perform any communications, only declare/retrieve data for/from a redistribution.

\begin{lstlisting}[float, caption=Redistribution functions for MaM., label=code:dmr-proteo-redist, captionpos=b]
real *x_, *u1_, *u2_, *u3_; // Data to redistribute

void create_slice_datatype(MPI_Datatype *slice) {
  int dimension_size = (M + 1) * (L + 1);
  MPI_Type_contiguous(dimension_size, mpi_real, slice);
  MPI_Type_commit(slice);
}
void mam_set() {
  real *redist_arrays[] = {x_, u1_, u2_, u3_};
  int halo_size = HALO * (M + 1) * (L + 1);
  for (int i=0; i<4; i++) {
    MAM_Data_add(redist_arrays[i] + halo_size, NULL, N, slice, MAM_DATA_DISTRIBUTED, MAM_DATA_VARIABLE);
  }
}
void mam_retrieve() {
  release();        // Release previous memory
  init();           // Allocate new memory
  real *aux_array;
  real *redist_arrays[] = {x_, u1_, u2_, u3_};
  int local_size = n * (M + 1) * (L + 1);
  int halo_size = HALO * (M + 1) * (L + 1);
  for (int i=0; i<4; i++) {
    MAM_Data_get_pointer(&(aux_array), i, NULL, NULL, MAM_DATA_DISTRIBUTED, MAM_DATA_VARIABLE);
    memcpy(redist_arrays[i] + halo_size, aux_array, size_real * local_size);
  }
}
\end{lstlisting}

In our original conference paper~\cite{iserte_towards_2025}, the malleable version of MPDATA supported resizing, but it was limited to process counts that were powers of two. This restriction simplified load balancing before malleability was implemented. However, to improve resource utilization, the load-balancing mechanism has been improved to allow any natural number of processes.

In addition, the $N$-dimension of the 3D grid (with dimensions $N$, $M$, and $L$) was evenly divided among the processes. As a result, the number of processes, \textit{SIZE}, needed to be a divisor of the grid size in the $N$-dimension, ensuring that each process owned an equal number of elements along this dimension. 
In the new approach, a block distribution along the $N$-dimension is also applied, considering the case where the size of the $N$-dimension is not a multiple of the number of processors, so that the difference between the maximum and the minimum of elements is one.
Listing~\ref{code:mpdata_load_balancing} illustrates the updated procedure for computing the local grid sizes. In this implementation, \textit{RANK} represents the rank of the process, and $N$, $M$, and $L$ correspond to the grid size dimensions.

This enhancement allows MPDATA to make full use of available resources and leverage all the features provided by MaM.

\begin{lstlisting}[float, caption=Optimized load balancing procedure for MPDATA allowing any natural number of processes., label=code:mpdata_load_balancing, captionpos=b]
void get_distribution() {
  int exp_qty = N / SIZE;
  int rem = N % SIZE;
  int ini, end;
  if(RANK < rem) {          // First subgroup
    ini = RANK * exp_qty + RANK;
    end = ini + (exp_qty + 1);
  } else {                  // Second subgroup
    ini = RANK * exp_qty + rem;
    end = ini + exp_qty;
  }
  n = end - ini;
  local_grid = (n + 2 * HALO) * (M + 1) * (L + 1);
}
\end{lstlisting}

\section{Experimental Results}\label{sec:experiments}
This section details the experimental platform and its configuration.
The experiments evaluate \textit{Baseline} and \textit{Merge} reconfiguration techniques and their asynchronous versions.

\subsection{Setup}\label{subsec:setup}
The evaluation is performed in the general-purpose partition of the Marenostrum 5 (MN5) supercomputer, a pre-exascale machine integrated into the EuroHPC-JU European supercomputing infrastructure.
The nodes of this partition are based on Intel Sapphire Rapids (4\textsuperscript{th} generation of Intel Xeon Scalable Processors), each one equipped with two Intel Xeon Platinum 8480+, with 56 cores each, running at 2~GHz, for a total of 112 cores and 256~GB of DDR5 memory.
The nodes are interconnected through a 100~Gbit/s ConnectX-7 NDR200 InfiniBand network.

Regarding the software stack, the executions rely on MPICH-4.2.1 and a customized version of \review{Slurm-17.02.0-0pre1} to support malleability included in the dynamic resource management framework DMRlib. In turn, DMRlib is configured with the MaM reconfiguration engine to enable process layout reshaping, as detailed in Section~\ref{subsec:reconf}.
Slurm is deployed using a node running the Slurm controller daemon, while the remaining nodes act as compute nodes. 
Slurm is configured with \texttt{sched/backfill}, \texttt{priority/multifactor}, and the custom  \texttt{select/natural} (introduced in Section~\ref{subsec:dynres}) plugins.

The experiments are scoped within 32 nodes for up to 2 hours, which fits in the QoS of the debug partition in MN5. Notice that a node from the allocation acts as the controller, while the remaining 31 nodes act as actual compute nodes.

\review{To deploy the experiments, DMRlib is launched within a standard Slurm allocation using an \texttt{sbatch} submission. Once the resources are granted, the nested malleable Slurm instance is deployed within the allocated environment. Interactions with either the parent or nested Slurm instance are seamlessly managed through the appropriate configuration of environment variables, allowing direct commands to the intended scheduler context.}

\begin{figure*}
    \centering
    \includegraphics[clip,width=0.999\linewidth,trim={0.5cm 0.1cm 4cm 0.1cm}]{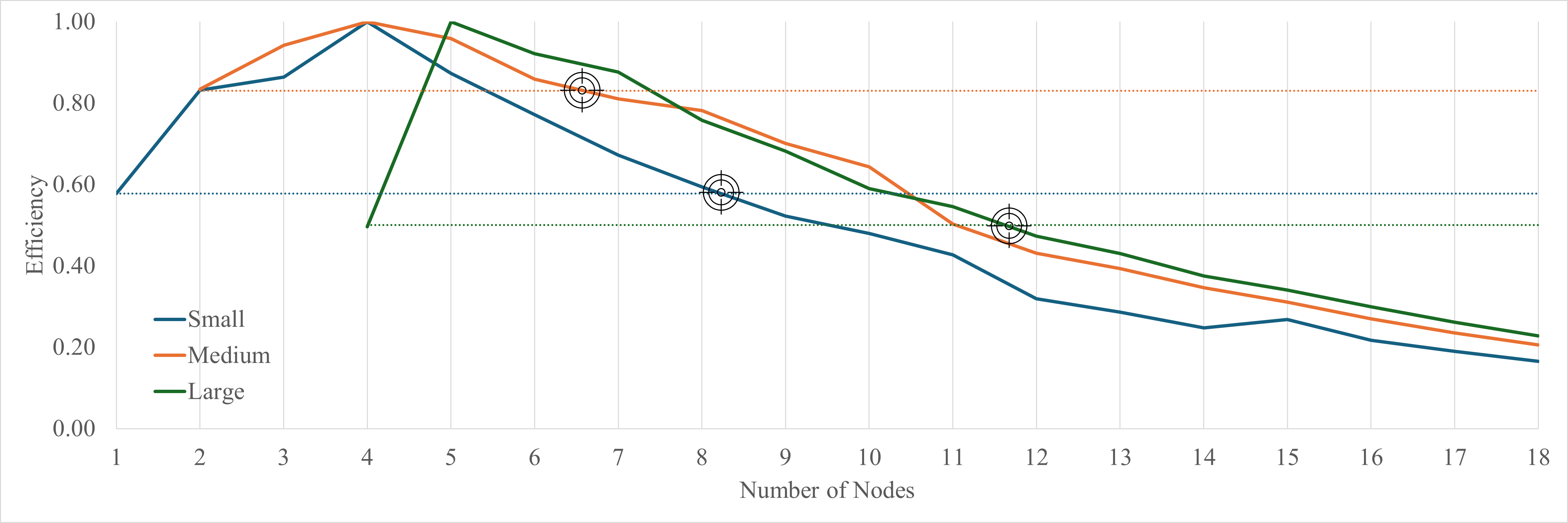}
    \caption{MPDATA normalized efficiency.}
    \label{fig:mpdata_scal}
\end{figure*}

Each experiment will execute a workload of 30 jobs instantiating MPDATA in three different flavors. In addition, each experiment is repeated five times to obtain more reliable statistics.
Table~\ref{tab:mpdata_conf} describes how jobs are classified depending on their problem size (PS) and the number of iterations.
\begin{table}[tbp]
    \centering
    \caption{MPDATA Configuration.}
    \label{tab:mpdata_conf}
    \begin{tabular}{rcl}
        \toprule
         \textbf{Type} & \textbf{Problem Size (PS)} & \textbf{Iterations} \\ \midrule
        Small  & 8,192 x 2,048 x 128  & 20  \\
        Medium & 8,192 x 4,096 x 128  & 60  \\
        Large  & 8,192 x 8,192 x 128  & 100 \\ \bottomrule
    \end{tabular}
\end{table}
In this regard, Table~\ref{tab:iter_times} showcases the 10-sample average iteration times for different quantities of nodes and job types.
\begin{table}[tbp]
    \centering
    \caption{MPDATA iteration times ($T_i$) for different number of nodes ($N_i$) and configuration types.}
    \label{tab:iter_times}
    \begin{tabular}{rccr}
        \toprule
        \textbf{Nodes ($N_i$)} & \textbf{Small} & \textbf{Medium} & \textbf{Large} \\ \midrule
        1  & 8.90 s.&  &  \\
        2  & 3.40 s.& 8.70 s.&  \\
        3  & 2.20 s.& 5.30 s.&  \\
        4  & 1.47 s.& 3.80 s.& 12.20 s.\\
        5  & 1.31 s.& 3.14 s.& 6.10 s.\\
        6  & 1.20 s.& 2.84 s.& 5.40 s.\\
        7  & 1.14 s.& 2.54 s.& 4.80 s.\\
        8  & 1.09 s.& 2.28 s.& 4.65 s.\\
        9  & 1.06 s.& 2.19 s.& 4.44 s.\\
        10 & 1.01 s.& 2.09 s.& 4.39 s.\\
        11 & 0.99 s.& 2.23 s.& 4.19 s.\\
        12 & 1.08 s.& 2.24 s.& 4.18 s.\\ \bottomrule
    \end{tabular}
\end{table}
This scalability analysis is performed by running a process per core, for a total of 112 ranks per node, corresponding to the 112 cores per node available in MN5.
The study reveals how times reach an asymptote as long as the iterations are run with more processes.
Particularly, Figure~\ref{fig:mpdata_scal} depicts how the different MPDATA configurations behave in terms of normalized efficiency (ranging from 0 to 1), which is calculated using the following equation:

\begin{equation}
    \text{Normalized Efficiency} = \frac{\text{Efficiency} - \text{Min}}{\text{Max} - \text{Min}},
\label{eq:normalized_efficiency}
\end{equation}

\noindent
where:
\begin{itemize}
    \item \(\text{Efficiency}\) is the actual value to be normalized,
    \item \(\text{Min}\) is the minimum efficiency in the series, and
    \item \(\text{Max}\) is the maximum efficiency in the series.
\end{itemize}

\noindent
The efficiency for a given configuration of nodes, denoted as $E_i$, is computed using:

\begin{equation}
    E_i = \frac{S_i}{{N_i}/{N_0}},
\label{eq:efficiency}
\end{equation}

\noindent
where:
\begin{itemize}
    \item \(N_0\) is the minimum number of nodes required to successfully run an instance of MPDATA for a given problem size,
    \item \(N_i\) is the number of nodes used in the configuration, and
    \item \(S_i\) is the speedup, calculated as $S_i = \frac{T_0}{T_i}$, being \(T_0\) the iteration time with $N_0$ nodes, and \(T_i\) is the iteration time with \(N_i\) nodes.
\end{itemize}

The same 30-job workload is leveraged in all the experiments.
The workload is presented in two flavors: static and dynamic; Depending whether its jobs are malleable.

Static jobs are submitted with the maximum number of nodes that provide an efficiency not lower than the efficiency achieved in \(N_0\).
Thanks to Figure~\ref{fig:mpdata_scal}, we can spot those maximums in the node configurations where the efficiency drops below \(N_0\) (in the figure those events are highlighted with a pointer).
Column ``Max. Nodes'' in Table~\ref{tab:jobs_conf} summarizes the number of nodes for each type of job.
Dynamic jobs are configured to work within a given range of nodes. While the upper limit matches the static configuration for each type, the lower limit corresponds to the \(N_0\) configuration, as column ``Min. Nodes'' in Table~\ref{tab:jobs_conf} contains.
Additionally, malleable jobs activate a DMRlib mechanism that prevents reconfigurations for four consecutive iterations.

\begin{table}[tbp]
    \centering
    \caption{Jobs Configuration.}
    \label{tab:jobs_conf}
    \begin{tabular}{rccc}
        \toprule
         \textbf{Type} & \textbf{Likelihood} & \textbf{Min. Nodes (\(N_0\)) } & \textbf{Max. Nodes}\\ \midrule
        Small  & 20\% & 1 & 8   \\
        Medium & 20\% & 2 & 6   \\
        Large  & 60\% & 4 & 11  \\ \bottomrule
    \end{tabular}
\end{table}

Jobs are included in the workload following a weighted random function honoring the likelihoods in Table~\ref{tab:jobs_conf} for each job type.
Furthermore, the Feitelson model~\cite{Feitelson1996}, configured with a mean interarrival time of $10$, is used to determine the delays among consecutive job submissions.

\subsection{Performance Analysis}
The workload is executed in two configurations: static and dynamic. For the dynamic execution, two reconfiguration techniques are evaluated, Baseline and Merge (see Section~\ref{subsec:background-proteo}).

\begin{figure}[tbp]
    \centering
    \includegraphics[clip,width=\linewidth, trim={0.1cm 0.7cm 1.8cm 1.0cm}]{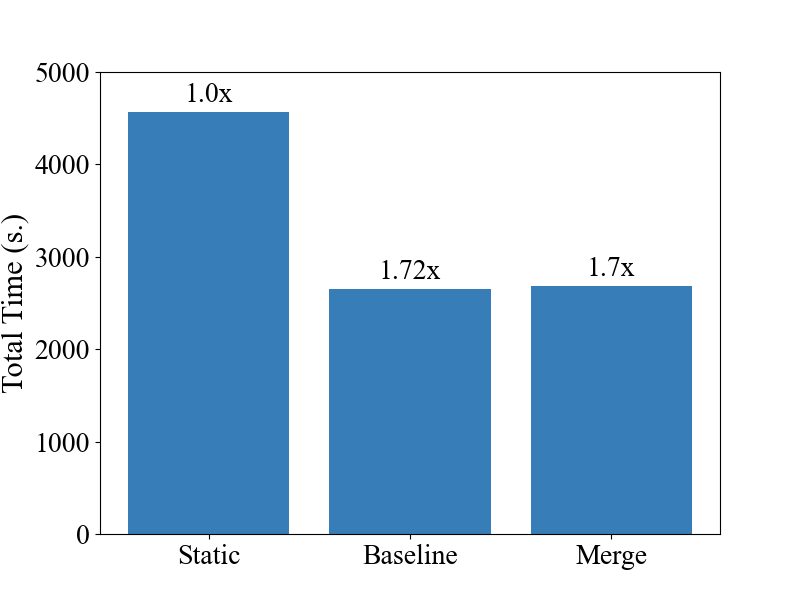}
    \caption{Workload Median Completion Times.}
    \label{fig:comp_times}
\end{figure}

Figure~\ref{fig:comp_times} presents the completion times of the three workload types in seconds. Each bar is labeled with its corresponding speedup relative to the first bar, which represents the workload executed without malleability. The second (Baseline) and third (Merge) bars correspond to dynamic workloads.

The results demonstrate a makespan reduction for dynamic workloads around 41\% compared to the static approach. Between dynamic workloads, no substantial performance difference is observed. This similarity is attributed to the negligible overhead of reconfigurations compared to overall execution time. A more detailed analysis of these results follows next.

\begin{figure}[tbp]
    \centering
    \begin{subfigure}[b]{0.99\columnwidth}
        \centering
        \includegraphics[width=\textwidth, trim={0.1cm 0.8cm 1.8cm 0cm}]{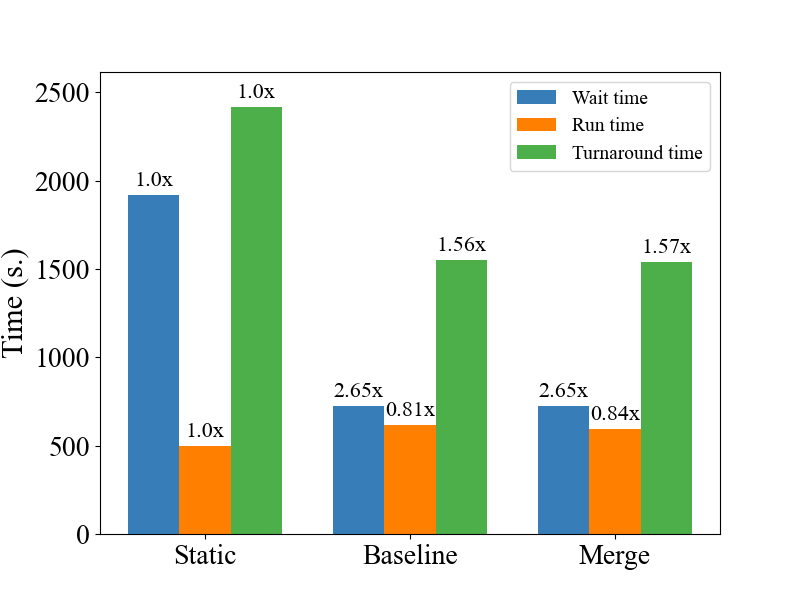}
        \caption{Large jobs.}   
        \label{fig:large}
    \end{subfigure}
    \\
    \begin{subfigure}[b]{0.99\columnwidth}
        \centering
        \includegraphics[width=\textwidth, trim={0.1cm 0.8cm 1.8cm 0cm}]{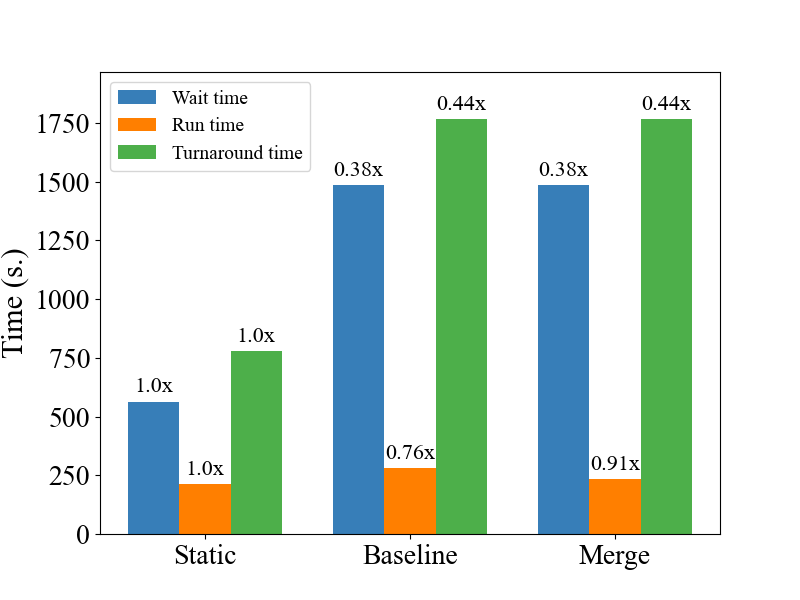}
        \caption{Medium jobs.}   
        \label{fig:medium}
    \end{subfigure}
    \\
    \begin{subfigure}[b]{0.99\columnwidth}
        \centering
        \includegraphics[width=\textwidth, trim={0.1cm 0.8cm 1.8cm 0cm}]{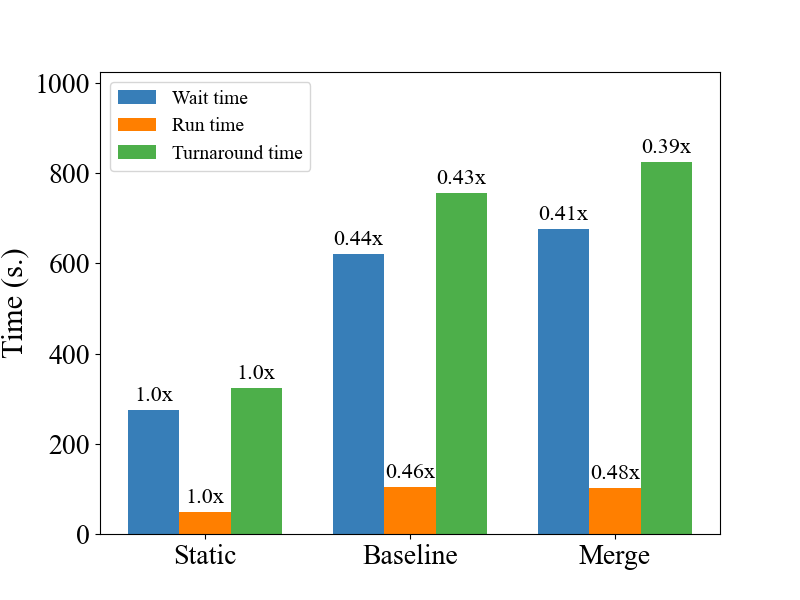}
        \caption{Small jobs.}   
        \label{fig:small}
    \end{subfigure}
    \caption{Median waiting, running, and turnaround times for jobs.}
    \label{fig:job_times}
\end{figure}

Figure~\ref{fig:job_times} depicts the median wait, run, and turnaround (sum of wait and run) times for the three job types (see tables~\ref{tab:mpdata_conf} and~\ref{tab:jobs_conf}) in each workload. 
The median is a representative metric when data are not normally distributed; besides, each bar is labeled with the corresponding speedup relative to the static workload times.

Among the job types, only large jobs significantly benefit from dynamic resource management, achieving a substantial reduction in wait time with a speedup of $2.65\times$, leading to a turnaround time speedup of around $1.5\times$.  
However, the run time increases because jobs are not always executed with the upper limit of resources, unlike in the static configuration.

\review{
Dynamic medium and small jobs are completed approximately $0.4\times$ slower because of the waiting time.
In that scenario, large jobs can be initiated with fewer resources, which reduces the available resources to backfill smaller jobs.
}
However, as shown in Figure~\ref{fig:comp_times}, the overall completion time improves, since large jobs dominate execution time, accounting for approximately $85\%$ of the workload time, compared to the medium and small jobs, contributing with $10\%$ and $5\%$ of the time, respectively. As a result, the turnaround time speedup in large jobs is the main factor driving the overall workload efficiency gains.

\begin{figure*}[tbp]
    \centering
    \includegraphics[clip,width=0.99\linewidth,trim={1.5cm 0.0cm 1.5cm 0.0cm}]{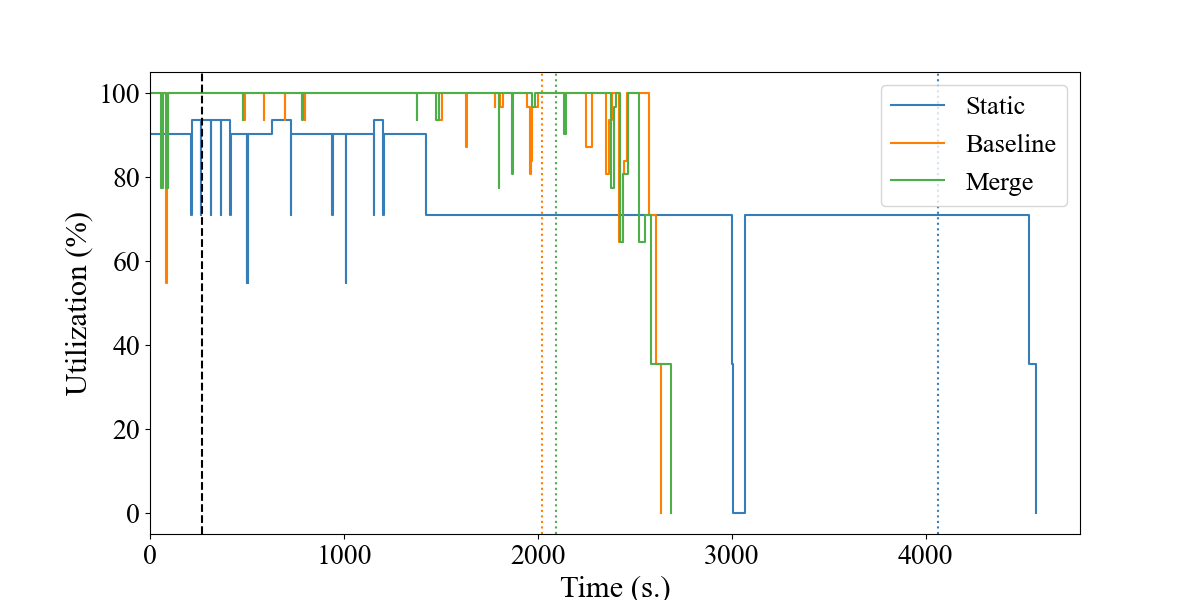}
    \caption{Median resource utilization rate for the different experiments.}
    \label{fig:plot_utilization}
\end{figure*}

Figure~\ref{fig:plot_utilization} illustrates the resource utilization rate for each workload type over its lifetime, with vertical dotted lines of the same color marking the point at which the waiting queue becomes empty for each experiment.
An additional vertical dashed black line is shown near the start of the execution and marks the point at which all jobs have been submitted.

Both dynamic approaches consistently achieve higher resource utilization than the static counterpart. 
Initially, utilization remains around $90\%$ in the static configuration, since small and medium jobs fill available idle resources. However, once these jobs are completed, utilization drops to $70\%$, since only large jobs remain, and no more than two can fit together. As a result, some nodes remain idle for the rest of the workload, leading to an overall mean utilization of $75.89\%$.

In contrast, the dynamic approaches maintain nearly $100\%$ utilization throughout their lifetime, except during brief moments of job resizes and terminations. These approaches achieve mean utilizations of $96.74\%$ for the Baseline method and $96.01\%$ for the Merge method.

In addition, the resource utilization rate shows why there is an increase in the wait time for medium and small jobs. In the first few minutes of all workloads, all jobs are submitted. In static workloads, all medium and small jobs begin execution between timestamps $0$ and $1150$, whereas in dynamic workloads, execution is distributed across the entire period. As a result, waiting times of medium and small jobs are reduced for the static workload.

\begin{figure}[tbp]
    \centering
    \includegraphics[clip,width=1\linewidth, trim={0.1cm 0.8cm 1.8cm 0cm}]
    {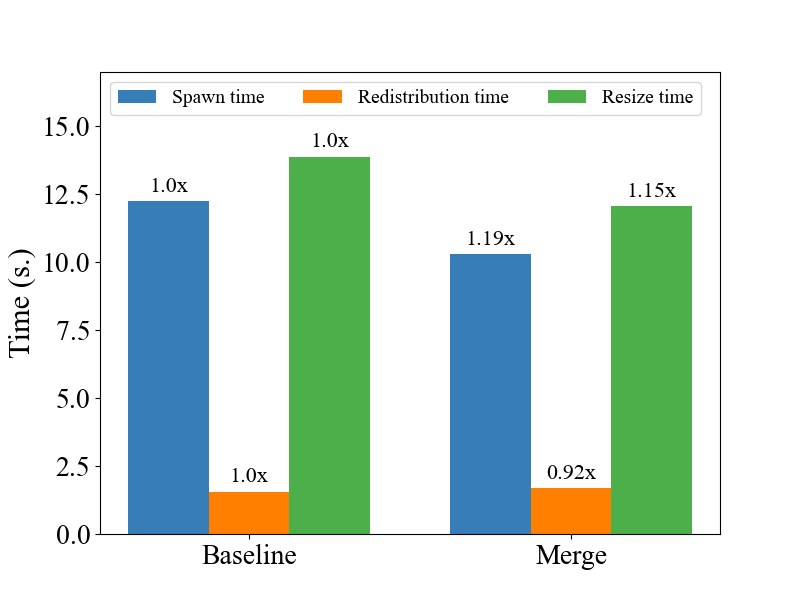}
    \caption{Median reconfiguration times.}
    \label{fig:resize_metrics}
\end{figure}

Figure~\ref{fig:resize_metrics} presents the medians of spawn time, data redistribution time, and total resize time for Baseline and Merge methods. Each bar counts with a label indicating the speedup relative to the Baseline method for the corresponding metric.

The spawn time is reduced in the Merge method, primarily due to the need to spawn fewer processes—or none at all during shrinkages. Conversely, the data redistribution time favors the Baseline method, yet, they have no significant statistical differences. To validate this observation, a Mann-Whitney U test was conducted to determine whether the medians of the two datasets are statistically different. This test accepted the null hypothesis of equal medians for the data redistribution time. 
The total resize time confirms that the Merge method is the optimal approach for resizing, achieving a speedup of $1.15\times$. The Mann-Whitney U test further supports this result by rejecting the null hypothesis of equal medians, indicating that the Merge method has a significantly smaller median resize time.

Despite the difference in resize time, the Baseline method slightly outperforms the Merge method in workload makespan and job turnaround time. This pattern has been consistent across the five experiments conducted for each workload type. Statistical tests applied to these metrics accepted the null hypothesis, showing no significant difference in medians.

In summary, although the Merge method reduces resize time, this improvement does not translate into a significant impact on overall workload metrics, because the time spent on resizing is relatively small.
Nevertheless, the Merge method’s ability to avoid oversubscription states makes it preferable to Baseline, since it needs less memory to reconfigure.
Specifically, $M_B \gg M_M$, where $M_B$ and $M_M$ represent the maximum memory required during a reconfiguration for Baseline and Merge, respectively.

Equation~\ref{eq:mem_base} defines the memory required for reconfiguring in the Baseline Method, where $PS$ denotes the problem size, and $NS$ and $NT$ are the number of source and target ranks, respectively, at the beginning and end of the reconfiguration process.

\begin{equation}
    M_B = \frac{PS}{NS} + \frac{PS}{NT} + C
\label{eq:mem_base}
\end{equation}

Similarly, Equation~\ref{eq:mem_merge} defines the memory required for reconfiguring using the Merge method.
It is important to note that, in both cases, there is a constant amount of memory ($C$) required by MaM to internally register part of the application data.

\begin{equation}
    M_M = \frac{PS}{min(NS, NT)} + C
\label{eq:mem_merge}
\end{equation}

More importantly, dynamic workloads nearly maximize resource utilization, reduce workload makespan, and improve the median job turnaround time. 
These benefits highlight the critical role of dynamic resource management techniques in enhancing productivity and efficiency.

\subsection{Asynchronous Reconfigurations}
The following analysis considers the usage of asynchronous reconfiguration strategies provided by MaM. With these approaches the application continues iterating while the resize is performed in the background. As such, the asynchronous counterparts of the dynamic workloads have been executed and studied, particularly the Baseline asynchronous (BaselineAsync) and the Merge asynchronous (MergeAsync).

Table~\ref{tab:workload_times} presents the median makespan and resource utilization rate for each experiment. Compared to the synchronous counterparts, the use of background resizing techniques increases the completion time, resulting in a speedup of approximately $0.95\times$, while reducing resource utilization by around $2\%$, regardless of the chosen method. However, a more in-depth analysis is needed to identify the specific factors contributing to the performance degradation observed with background resizing.

\begin{table}[tbp]
    \centering
    \caption{Median workload metrics.}
    \label{tab:workload_times}
    \begin{tabular}{rccl}
        \toprule
         \textbf{Experiment} & \textbf{Makespan} & \textbf{Utilization} \\ \midrule
        Static  & 4,567 s. & 75.89\%   \\
        Baseline & 2,655 s. & 96.74\%   \\
        Merge  & 2,685 s. & 96.01\%    \\
        BaselineAsync & 2,778 s. & 94.60\%    \\
        MergeAsync & 2,790 s.& 95.48\%   \\ \bottomrule
    \end{tabular}
\end{table}

\begin{figure}[tbp]
    \centering
    \begin{subfigure}[b]{0.99\columnwidth}
        \centering
        \includegraphics[clip,width=\textwidth,trim={0.1cm 0.8cm 1.8cm 0cm}]
        {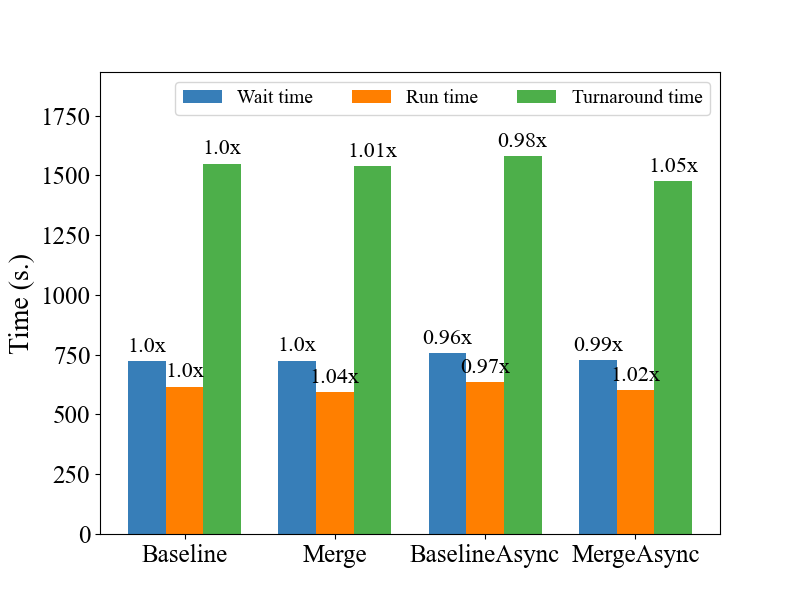}
        \caption{Large jobs.}   
        \label{fig:large_A}
    \end{subfigure}
    \\
    \begin{subfigure}[b]{0.99\columnwidth}
        \centering
        \includegraphics[clip,width=\textwidth,trim={0.1cm 0.8cm 1.8cm 0cm}]
        {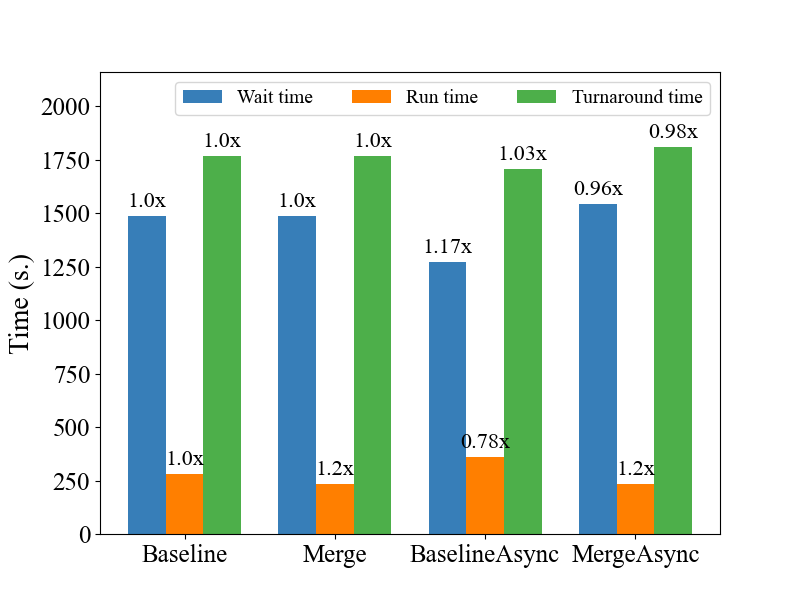}
        \caption{Medium jobs.}   
        \label{fig:medium_A}
    \end{subfigure}
    \\
    \begin{subfigure}[b]{0.99\columnwidth}
        \centering
        \includegraphics[clip,width=\textwidth, trim={0.1cm 0.8cm 1.8cm 0cm}]
        {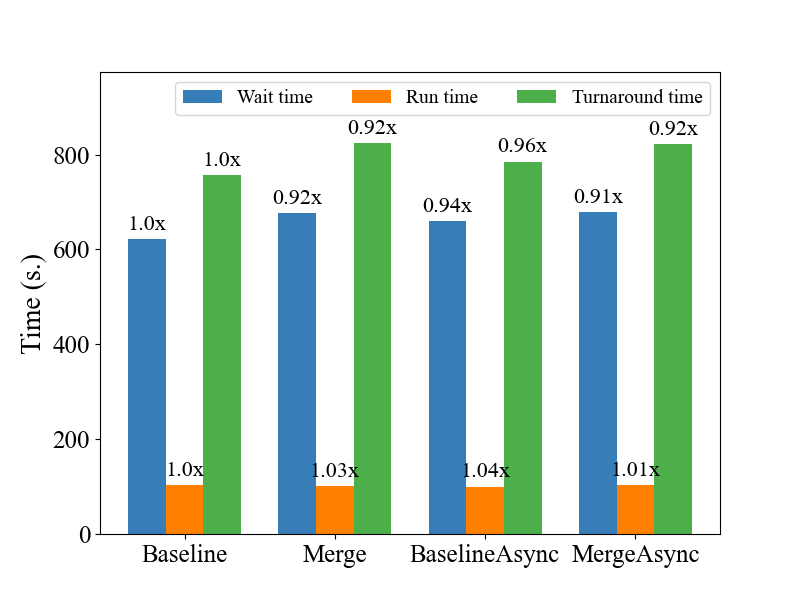}
        \caption{Small jobs.}   
        \label{fig:small_A}
    \end{subfigure}
    \caption{Median waiting, running, and turnaround times for jobs in the dynamic workloads.}
    \label{fig:job_metrics_A}
\end{figure}

Figure~\ref{fig:job_metrics_A} shows the median wait, run, and turnaround times for the three different job types in each experiment. Each bar includes a label indicating the speedup relative to Baseline synchronous.

When tackling large jobs, the choice of reconfiguration method does not significantly impact performance. 
The observed differences in runtime are primarily due to the inherent variability of the execution operation itself, not the chosen strategy. 
However, for medium jobs, some differences emerge. 
The Merge methods achieve a speedup in runtime of $1.2\times$, but since turnaround time is monopolized by the waiting time, this does not significantly affect turnaround time.
This improvement is mainly due to the Merge method preventing oversubscription effects during reconfigurations. 
The BaselineAsync method demonstrates a more substantial degradation by reducing its jobs run time to $0.78\times$.

For small jobs, all methods perform worse compared to the synchronous Baseline method. 
Their wait time speedup decreases to approximately $0.9\times$, with only a slight increase in the run time. 
Since wait time dominates the turnaround time for small jobs, all methods result in longer overall turnaround times.

These slight increments in the jobs turnaround times align with the results in Table~\ref{tab:workload_times}, which show speedups ranging from $0.95\times$ to $0.98\times$ for Merge, BaselineAsync, and MergeAsync methods.

\begin{figure}[tbp]
    \centering
    \includegraphics[clip,width=0.99\linewidth, trim={0.1cm 0.8cm 1.8cm 0cm}]{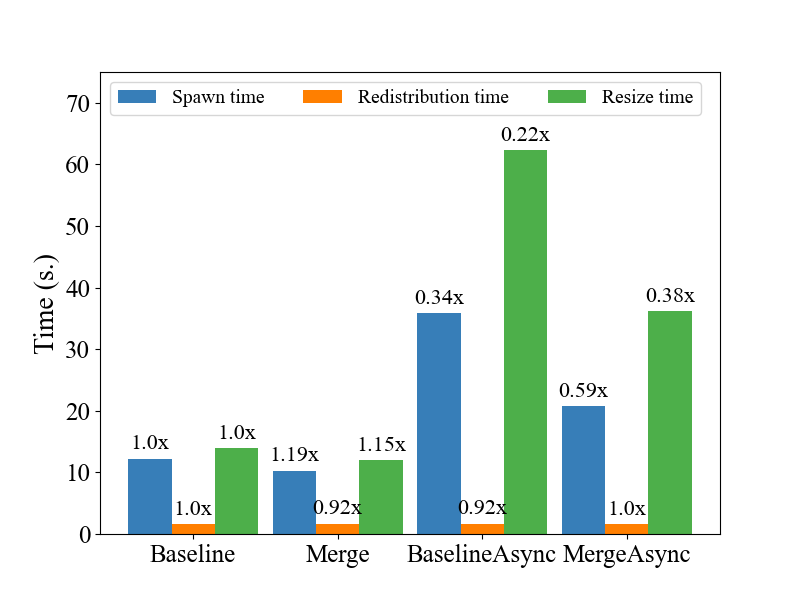}
    \caption{Median reconfiguration times with asynchronous experiments.}
    \label{fig:resize_metrics_A}
\end{figure}

Figure~\ref{fig:resize_metrics_A} presents the median times of process spawning, data redistribution, and total resize, for all the dynamic experiments. Each bar includes a label indicating the speedup relative to the Baseline method for the corresponding metric.

In both asynchronous methods, the speedup is primarily reduced due to the spawn time, which decreases by $0.34\times$ for BaselineAsync and $0.59\times$ for MergeAsync. 
This slowdown occurs because spawning processes in the background require creating a separate thread for each rank. 
However, data redistribution time remains unaffected, since it relies on MPI’s non-blocking functions.

As a result, the total resize time increases, with speedups of $0.22\times$ for BaselineAsync and $0.38\times$ for MergeAsync. 
These findings indicate that asynchronous approaches are not beneficial for reducing resize times. 
Moreover, they may negatively impact jobs run times, ultimately affecting makespan times.

Table~\ref{tab:resize_times} provides the median values for the accumulated resize time over the workload, the number of iterations performed during asynchronous reconfigurations, and the speedup of those iterations. 
The iteration speedup ($IS$), calculated as $IS = \frac{T_o}{T_a}$, compares the iteration time ($T_o$) for a given node configuration (see Table~\ref{tab:iter_times}) with its asynchronous iteration time ($T_a$) counterpart.

\begin{table}[tbp]
    \centering
    \caption{Median reconfiguration metrics.}
    \label{tab:resize_times}
    \begin{tabular}{rccc}
        \toprule
         \textbf{Experiment} & \textbf{Accumulated} & \textbf{Overlapped} & \textbf{Speedup} \\
          & \textbf{Resize Time} & \textbf{Iterations} & \textbf{($IS$)}\\ \midrule
        Baseline & 189.43 s. & 0 & $1\times$\\
        Merge & 112.88 s. & 0 & $1\times$\\
        BaselineAsync & 678.73 s. & 197 & $0.41\times$ \\
        MergeAsync & 428.78 s. & 187 & $0.60\times$ \\ \bottomrule
    \end{tabular}
\end{table}

Synchronous reconfigurations accounts for approximately $5\%$ to $7\%$ of the total workload makespan. This overhead is negligible compared to the overall reduction in execution time achieved through dynamic resource utilization.

In contrast, for asynchronous reconfigurations, resizing overhead lies between $15\%$ and $25\%$ of the workload makespan. This increase is due to iterations continuing while a resize occurs in the background. 
However, during this period, around 200 iterations are executed, albeit at a reduced speedup of approximately $0.5\times$, meaning each iteration takes twice as long.

The primary reason asynchronous methods are not preferred is this slowdown, combined with the fact that most resizes involve expanding the job. When increasing resources, the new iteration time will naturally be shorter, making resizing as soon as possible more beneficial than performing additional iterations with a reduced speedup.

\section{Applicability}\label{sec:app}
This work has focused on a single application (MPDATA) to generate the workload for our evaluation. However, the results are consistent with prior studies involving DMR and other malleability frameworks~\cite{aliaga_survey_2022}. For example, one of the most comprehensive evaluations~\cite{Iserte2020} included a mix of four different applications, generating both homogeneous and heterogeneous workloads with varying proportions of dynamic jobs. Drawing from that work and the outcomes presented in this paper, we can extrapolate that dynamic resource management proves especially beneficial for workloads composed of malleable jobs with limited scalability. In contrast, applications that exhibit near-perfect scalability may not benefit significantly from malleability, as they already make efficient use of resources. 

In terms of general applicability to other codes beyond MPDATA, the DMR framework provides a structured logic centered on initialization, reconfiguration, and finalization phases. This approach allows many existing HPC applications to be adapted for malleability, as demonstrated in previous work~\cite{Iserte2020, Iserte2019a}. Applications must be able to define a clear initialization phase that can be executed at launch and after each reconfiguration, a reconfiguration phase to redistribute data, and a finalization step to properly release memory. While certain large-scale codes cannot be made malleable out of the box, the modularity and abstractions provided by DMR significantly reduce the porting effort. Ongoing work within the DMR team is currently focused on extending malleability support to such complex applications.

It is also worth noting that the proposed resource selection policy, \texttt{select/natural}, developed for the Slurm plugin system, operates independently of existing job scheduling and prioritization plugins such as \texttt{sched/backfill} and \texttt{priority/multifactor}. Establishing coordination between these components could pave the way for holistic scheduling strategies that better integrate dynamic jobs into the global queue, optimizing both resource assignment and job throughput.
This is a worthwhile branch to explore in the field of scheduling.

Finally, the current work and its evaluation rely on MPICH-4.1, the MPI distribution supported by Proteo’s malleability engine. While most MPI libraries comply with the MPI standard~\cite{mpi_forum_mpi_2015}, their internal implementations vary, particularly in how certain operations are handled. Therefore, extending MaM to support other MPI implementations, such as Open MPI, is a natural and necessary future direction to broaden its applicability. DMR will seamlessly integrate MaM versions.

\section{Conclusions} \label{sec:conclusions}
This work addresses the need for a dynamic resource manager that adheres to the standard MPI, supports multiple reconfiguration strategies, integrates seamlessly with a widely used HPC resource management system such as Slurm, and remains easy to use for scientific application developers. To meet these requirements, we have developed and evaluated DMR as a production-ready dynamic resource management solution that enhances flexibility, adaptability, and efficiency in modern HPC environments.  

A review of the state of the art reveals a lack of a transversal dynamic resource manager that spans both the parallel runtime and the resource manager while relying on standard MPI implementations and widely adopted RMS frameworks. DMR fills this gap by enabling standardized dynamic resource management and democratizing its adoption. Through its integration with Slurm and support for standard MPI, DMR provides a unified approach that simplifies malleability and resource elasticity in HPC workloads.  

To showcase its capabilities, we implemented the MaM interfaces for DMR and developed a malleable version of the MPDATA application. Additionally, we established a synergy between MaM and DMRlib, allowing seamless support for Slurm while incorporating advanced spawning strategies. These integrations demonstrate the versatility of DMR in supporting a wide range of reconfiguration methods while remaining transparent and user-friendly.  

Experimental results confirm that our dynamic resource management approach significantly improves efficiency with minimal programming effort. Our findings indicate that enabling dynamic resource allocation allows a workload to be completed in just 60\% of the time required by a static configuration while increasing over $20\%$ the resource utilization. This highlights the practical benefits of dynamic malleability, reinforcing its value for HPC workloads.  

Furthermore, this work also evaluates additional malleability strategies provided by MaM, including merging MPI communicators and asynchronous reconfigurations, which enhance the versatility of dynamic workloads. 
And, although these new strategies do not yield a significant performance improvement in the evaluated workloads, they introduce new opportunities by avoiding the need to respawn all MPI ranks, thereby reducing memory consumption, as well as performing reconfigurations in the background when needed.

With these advancements, we hope the DMR methodology to seed technology for integrating malleability into HPC environments, enabling smarter and more adaptive resource management strategies for next-generation supercomputing.  


\section*{CRediT authorship contribution statement}
\begin{itemize}
\item Sergio Iserte: Conceptualization, Methodology, Investigation, Resources, Writing - Original Draft, Project administration.
\item Iker Martín-Álvarez: Methodology, Software, Investigation, Validation, Formal analysis, Data Curation, Writing - Original Draft.
\item Krzysztof Rojek: Formal analysis, Supervision.
\item José I. Aliaga: Software, Writing - Review \& Editing, Supervision, Funding acquisition.
\item Maribel Castillo: Software, Formal analysis, Writing - Review \& Editing, Supervision.
\item Weronika Folwarska: Software.
\item Antonio J. Peña: Writing - Review \& Editing, Supervision, Funding acquisition.
\end{itemize}

\section*{Declaration of competing interest}
The authors have no conflicts of interest to declare that are relevant to the content of this article.

\section*{Funding sources}
The researchers from BSC are involved in The European PILOT project, which has received funding from the European High-Performance Computing Joint Undertaking (JU) under grant agreements No. 101034126 and No. PCI2021-122090-2A funded by MICIU\slash AEI\slash 10.13039\slash 501100011033 and EU NextGenerationEU\slash PRTR.
They are also grateful for the support from the Department of Research and Universities of the Government of Catalonia to the AccMem (Code: 2021 SGR 00807).
Antonio J. Peña was partially supported by the Ramón y Cajal fellowship RYC2020-030054-I funded by MCIN\slash AEI\slash 10.13039\slash 501100011033 and by ``ESF Investing in your future''.

The researchers from UJI have been funded by the project PID2023-146569NB-C22 supported by MICIU\slash AEI\slash 10.13039\slash 501100011033 and ERDF/UE.
Researcher I.~Martín-Álvarez was supported by the predoctoral fellowship ACIF\slash 2021\slash 260 from Valencian Region Government and European Social Funds.

\section*{Data availability}
The source codes for DMR (\url{https://gitlab.bsc.es/siserte/dmr}), MaM (\url{https://lorca.act.uji.es/gitlab/martini/malleability_benchmark/-/tree/PPAM24-JournalSpecialIssue}) and MPDATA (\url{https://gitlab.bsc.es/siserte/mpdata-dmr}) are publicly available. All data used to create the tables and figures in the Experimental Results section are available at \url{https://doi.org/10.5281/zenodo.14812022}. 

\bibliographystyle{unsrt}
\bibliography{bib}

\end{document}